\documentclass[journal]{IEEEtran}

\usepackage{cite}

\ifCLASSINFOpdf
\else
\fi

\usepackage{amsmath,amssymb}
\usepackage{color}
\usepackage[caption=false,font=footnotesize]{subfig}
\usepackage{lipsum,booktabs,multirow}
\usepackage{url}
\usepackage{amsmath}
\usepackage{graphicx}

\makeatletter
\def\input@path{{./tables/}}
\makeatother
\graphicspath{{./figures/}}

\newcommand{\nnz}{\mathrm{nnz}}

\newcommand{\Smp}{S_{\mathrm{MP}}}

\newcommand{\EC}{\mathrm{EC}}
\newcommand{\MEC}{\mathrm{MEC}}

\newcommand{\PlssC}{P_{\mathrm{Loss}}^{\mathrm{Conv.}}}
\newcommand{\PlssN}{P_{\mathrm{Loss}}^{\mathrm{Ntwk.}}}
\newcommand{\Pdc}{P_{\mathrm{dc}} }
\newcommand{\ScTot}{S_{c}^{\mathrm{Total}} }

\begin{document}

\title{Defining and Constraining the Electrical Cardinality\\of Multiport Converter Mission Profiles}

\author{Matthew~Deakin,~\IEEEmembership{Member,~IEEE}
\thanks{This work was supported by the the Royal Academy of Engineering under the Research Fellowship programme, and the EPSRC Multi-energy Control of Cyber-Physical Urban Energy Systems (MC2) project (EP/T021969/1). M. Deakin is with Newcastle University, Newcastle-upon-Tyne, UK. Contact: \texttt{matthew.deakin@newcastle.ac.uk}.}
}

\maketitle

\begin{abstract}
Mission profiles describe a representative set of conditions that a power converter is designed to operate under, and are known to be more complicated for multiport converter applications due to a wider range of combinations of powers that can be transferred between ports. This paper studies the properties of mission profiles derived from operational optimization of multiport converters in distribution system applications (e.g., soft open points). The electrical cardinality of the mission profile is introduced as a useful, naturally varying property of multiport mission profiles derived from optimal operation within distribution system, with the cardinality equal to the number of non-zero power transfers at a given time. Furthermore, it is shown that the cardinality can be conveniently constrained within the framework of conventional mixed-integer conic optimization problems, yielding a family of mission profiles that can enable converter designers to simplify multiport designs via converter reconfiguration. Results demonstrate the potential to reduce the cardinality in a four-terminal multiport converter from four to two whilst still effectively supporting congestion management and achieving 91.7\% of the loss reduction capabilities of a conventional design.
\end{abstract}


\IEEEpeerreviewmaketitle

\section{Introduction}

\IEEEPARstart{A}{pplications} of power electronics-based power converters in power systems today are rapidly growing, with industry requiring new methods of integrating electric transportation and renewable generation into the ac grid.  Multiport converters are an important class of grid-connected power converters, allowing power to be efficiently transferred between different feeders in a power distribution system. For example, a three terminal multiport converter can act as a soft open point (SOP), connecting two ac distribution feeders through a dc link with an integrated energy storage system that can provide temporal flexibility.

A mission profile is a representative time series collecting the parameters that affect the key performance indices for a given power converter (e.g., reliability, lifetime or efficiency). This typically will consist of at least the thermal or electrical stresses that the converter will be under, but can also include other relevant environmental parameters such as humidity or vibration. Accurate mission profile simulation can be complex and time consuming, and so there has been significant research on this topic. For example, recent works focus on fast emulation of thermal and electrical aspects of mission profiles including realistic switching profiles \cite{wang2019viable}, or considering comprehensive sets of parameters, such as switching frequency or the direction of real power transfer for increasing accuracy \cite{zhan2022multi}.

The electric power distribution community is also concerned with the development of mission profiles, but typically from the point of deriving optimal power transfers for a given converter design and optimization approach, rather than simulating the impacts on long-term converter reliability or state of health. For example, a three-terminal ac/dc/ac multiport, operated as an integrated SOP-energy storage device, is scheduled to provide loss reduction, price arbitrage and robust congestion management in \cite{sarantakos2022robust}, with the main focus of the work exploring impacts of uncertainty on the feeder voltages. Similarly, a robust strategy is developed for a five-terminal SOP in \cite{sun2020optimized}, with the main contribution a two-stage optimization combining both semidefinite programming and droop control, again considered over a representative day.

Of those works that do consider properties of mission profiles that result from distribution system optimization, the metrics considered typically do not clearly link to potential impacts on converter design. For example, average and time-varying utilization of power transferred through individual converters are presented to summarise performance in \cite{lou2020new} for a two-terminal phase changing SOP. Similarly, in \cite{deakin2022design}, the utilization of individual power converter's mission profiles are studied as a function of variable converter sizes. The variability of the electrical parameters of the mission profile are shown to influence efficiency in \cite{schaumburg2023efficiency}, proposing redeployment of modular converters within a multiport converter. The dimension and volume of a multiport's capability chart is studied in \cite{deakin2023multiplexing_inpress}, with it recognized that multiport converters with degenerate capability charts (i.e., with a capability chart hypervolume of zero) can still yield good performance within a network. Indeed, this final observation clearly motivates the study of cardinality of the power transfers within mission profiles to explore and understand how prevalent these simpler designs might be in future \cite{deakin2023multiplexing_inpress,majumder2021distribution}.

The contribution of this paper is to introduce the electrical cardinality of a mission profile, providing an intuitive property that can be used to study interactions between a distribution network's requirements and the design of reconfigurable multiport converters. The proposed definition is particularly useful as it conveniently links to cardinality constraints that can be imposed in mixed-integer conic optimization problems commonly used in distribution optimal power flow problems. The cardinality is subsequently used to explore the properties of optimal mission profiles of a given dimension, with case studies demonstrating a halving of the cardinality whilst still achieving 91.7\% of the potential system benefits.

The paper is structured as follows. In Section~\ref{s:method}, the electrical cardinality is introduced, with a short discussion highlighting why this naturally varies with the outputs of optimal distribution network scheduling, then showing how reconfigurable power converters can exploit mission profiles with reduced cardinality. Subsequently, cardinality constraints are proposed in Section~\ref{s:system_modelling} to demonstrate how a family of mission profiles can be developed that could allow a converter designer to explore a range of potential reconfigurable designs with varying complexity. Two case studies are presented and solved in Section~\ref{s:results} to demonstrate qualitative and quantitative attributes of mission profiles with varying cardinality constraints. Conclusions are drawn in Section~\ref{s:conclusions}.

\section{Mission Profile Electrical Cardinality}\label{s:method}

The cardinality of a vector is the number of non-zero elements of that vector. In this section, the electrical cardinality of a mission profile is introduced from this concept, with it demonstrated how it can be determined from the powers that are transferred to and from the ac distribution grid. Reconfigurable multiplexed soft open points (MOPs) are then introduced as a specific multiport topology that can take advantage of reduced cardinality, highlighting a direct application of the electrical cardinality.

\subsection{Defining Electrical Cardinality}

In this work, the electrical cardinality of a mission profile is defined as the cardinality of the powers transferred by a multiport converter into an ac distribution grid at a given time. It is referred to as the electrical cardinality so as to distinguish from the thermal aspects of the mission profile which are also of significant interest (but are not considered further in this work). For example, consider apparent powers transferred by an $ m $-terminal SOP over duration $ \tau $ collected in a matrix $ \Smp \in \mathbb{R}^{\tau \times m}$. In this case, the electrical cardinality ($ \EC $) at a given time instant $ \tau $ is defined as
\begin{equation}\label{e:electric_cardinality}
\EC[\tau] = \nnz(\Smp[\tau,:])\,,
\end{equation}
where $ \nnz $ returns the number of non-zero elements of a vector, with matrix $ \Smp $ indexed using `Matlab' notation. The EC can then be used to also define the maximum electrical cardinality ($ \MEC $) of the mission profile as
\begin{equation}\label{e:mec_defn}
\MEC = \max _{\tau} \, \EC[\tau]\,.
\end{equation}
When the EC needs to be calculated in practise (e.g., from a mission profile $ \Smp $ returned by a numerical optimization routine), it can be calculated based on a tolerance $ \epsilon $, e.g.,
\begin{equation}\label{e:nnz_tol}
\EC[\tau] = \sum _{i=1}^{m} I_{\epsilon}(\Smp[\tau,i])\,,
\end{equation}
where the indicator function $ I_{\epsilon} $ returns 1 if its argument is greater than $ \epsilon $, and zero otherwise. A relative value of $ \epsilon $ of $ 10^{-5} $ times the total converter capacity is used in this work.

There are several reasons that the value of the power transferred through a given terminal of a multiport converter might have a value of zero and therefore yield time-varying electrical cardinality $ \EC $. If a terminal is connected to a variable generator (such as solar) then the generator might have zero output for significant time periods. A distribution feeder might also be out of service. If a multiport is used most of the time for system loss reduction and arbitrage, the losses within the converter might be sufficiently high that it is not cost-effective to run the converter.

In fact, converter losses can tend to produce optimal solutions that are sparse in the apparent power transfers $ S_{c} $. For example, for converter losses of an $ m $-terminal multiport $ \PlssC \in \mathbb{R}^{m} $, loss coefficient $ k $ and linear model \cite{jiang2022overview,deakin2023multiplexing_inpress}
\begin{equation}\label{e:conv_loss}
\PlssC [i] = k S_{c}[i] \; \forall \; i \in  [1,\,m]\,,
\end{equation}
the solution of optimization problems that involve minimization of total converter losses will tend to result in a solution which is sparse in $ S_{c} $ (i.e., has few non-zero entries). This is because \eqref{e:conv_loss} acts as a `shrinkage operator' on $ S_{c} $ \cite{dall2014sparsity}. Increasing values of $ k $ will yield increasingly sparse solutions in $ S_{c} $ \cite[Ch. 6.2]{boyd2009convex}.

\subsection{Exploiting Partial Electrical Cardinality}

The electrical cardinality $ \EC $ can take integer values between 0 and $ m $ (for a multiport connected to $ m $ ac feeders), with cases that are of most interest in this work being when the $ \MEC $ is less than $ m $. This is because an $ \MEC $ less than $ m $ implies that a multiport converter with $ m $ legs has a degree of redundancy, insofar as there is potential for the exploitation of reconfigurable multiplexed soft open point (MOP) designs with a reduced number of ac/dc legs \cite{deakin2023multiplexing_inpress}. 

\begin{figure}\centering
\subfloat[Conventional SOP with integrated DG ]{\includegraphics[height=5.5em]{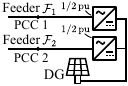}\label{f:reconfig_design_sop}}
~
\subfloat[Dual converter MOP (variable $ \alpha $, with $ 0< \alpha < 1$)]{\includegraphics[height=5.5em]{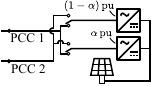}\label{f:reconfig_design_complex}}
~
\subfloat[Single converter MOP]{\includegraphics[height=5.5em]{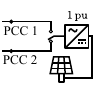}\label{f:reconfig_design_simple}~}

\caption{Single line diagram of a conventional SOP (with integrated DG connected to the DC link), (a), as compared to reconfigurable multiplexed SOP (MOP) multiport designs with two converters (b), or a single reconfigurable ac/dc converter (c). All designs have the same pu ac/dc converter capacity, but different capability charts and varying complexity of construction in terms of the number and type of components.}
\label{f:reconfig_design}
\end{figure}

For example, three multiport designs are plotted in Fig.~\ref{f:reconfig_design}, both with and without MOP reconfiguration stages. If the maximum cardinality $ \MEC $ has value two (i.e., it is required that at some time periods power must be transferred to or from point of common coupling (PCC) 1 to PCC~2 simultaneously), then the conventional design (Fig.~\ref{f:reconfig_design_sop}) or dual converter MOP design (Fig.~\ref{f:reconfig_design_complex}) can be considered. In contrast, if the $ \MEC $ has value unity, such that the solar power is exported only through PCC~1 or PCC~2 (or reactive power is never required for voltage control concurrently in both feeders), then the simpler MOP design Fig.~\ref{f:reconfig_design_simple} can also be considered. Such a design only requires a single ac/dc converter and can inject a full 1~pu into either PCC~1 and PCC~2, and so has advantages over both the conventional SOP and dual converter MOP in at least one sense.

\section{System Modelling}\label{s:system_modelling}

In the previous section, the electrical cardinality and its maximum value $ \MEC $ were introduced as metrics that can be used to explore the potential degree of redundancy of non-reconfigurable multiport converters. In this section, we first consider an additional benefit of defining the electrical cardinality: this cardinality can be varied in a computationally efficient way, enabling a multiport converter designer to explore the performance of a family of mission profiles. Cardinality constraints are introduced and then formulated using a big-$ M $ procedure, so that these constraints can be appended to existing convex optimization problems. Subsequently, a mixed-integer second order cone program is presented that integrates these cardinality constraints into an examplar optimal power flow problem for a distribution system, utilizing idealised MOP designs to bound the performance of all profiles of a given cardinality.

\subsection{Cardinality Constraints Formulated Using the Big-M Method}\label{ss:cardinality}

For an optimization decision variable $ y\in \mathbb{R}^{p} $ for some integer $ p $, a cardinality constraint \cite{mosek2021cookbook}
\begin{equation}\label{e:nnz_constraint}
\nnz(y) \leq q\,,
\end{equation}
for some $ q < p $ can be recast using a big-$ M $ formulation (this reformulation is required as the constraint as-written in \eqref{e:nnz_constraint} cannot be incorporated into integer optimization packages). Specifically, if $ y \geq 0 $ and $ M $ is an upper bound for each element of $ y $, then $ p $ auxiliary binary variables $ z_{i}\in \mathbb{B} $ can be introduced alongside the constraint
\begin{equation}\label{e:bigM}
y[i] \leq M z_{i} \; \forall \; i\,,
\end{equation}
such that if the $ i $th element of $ y $ is zero then so will the value of $ z_{i} $; if the $ i $th element of $ y $ is non-zero, then $ z_{i} $ will have value one. These auxiliary variables allow for \eqref{e:nnz_constraint} to then be rewritten
\begin{equation}\label{e:nnz_actual}
\sum _{i=1}^{p} z_{i} \leq q\,.
\end{equation}
Therefore, a optimization problem with constraints \eqref{e:bigM}, \eqref{e:nnz_actual} is equivalent to having constraint \eqref{e:nnz_constraint}. These equations are linear in binary variables $ z_{i} $, and so can be included directly in conventional mixed-integer conic optimization packages.

\subsection{System model}

For a reconfigurable MOP connected to an $ m $-feeder node within a distribution network, the goal is to schedule the real and reactive power flows transferred $ P_{c}\in \mathbb{R}^{m},\,Q_{c} \in \mathbb{R}^{m} $ to minimize network and converter losses, subject to congestion constraints. The optimization developed to solve this problem builds on the approach outlined in \cite{deakin2023multiplexing_inpress}, with a number of substantive changes as follows.
\begin{itemize}
\item The primary topic of this work is the exploration of the solution as the cardinality varies, and so a cardinality constraint of the form \eqref{e:nnz_constraint} is added and subsequently solved for a range of cardinality values $ q $ up to the number of terminals $ m $.
\item Instead of considering large numbers of variable MOP or SOP designs for a given converter capacity, instead idealised, fully reconfigurable designs are considered within the modelling \cite{deakin2023multiplexing_inpress}. This provides a convenient upper bound for the performance of a design with a given cardinality, although it is possible the mission profiles may not be realisable as designs (this is discussed further in Section~\ref{ss:discussion}).
\item To model network congestion, a linearized model is used that maps power injections to changes in voltage magnitudes. The `First Order Taylor' method is used for linearization from \cite{bernstein2018load}, linearized at the no-load solution.
\item Finally, the model of losses in power injections is based only on a quadratic model developed on the linearization of complex voltages \cite{bernstein2018load} around the no-load point (rather than calculating a linearization at all loading conditions separately).
\end{itemize}

With these changes, the optimization problem can be written as follows.
\begin{align}
\min _{P_{c},\,Q_{c}} & \PlssN + \sum_{i} \PlssC [i]  \label{e:opt_obj} \\
\mathrm{s.t.}\: S_{c}[i] & \geq \sqrt{P_{c}^{2}[i] + Q_{c}^{2}[i]} \: \forall \: i \in [1,\,n]\,, \label{e:opt_apparent_limit} \\
\sum \Pdc & = 0 \,, \label{e:opt_dc_kcl} \\
\Pdc [i] + \PlssC [i] & = P_{c}[i] \: \forall \: i \in [1,\,m]\,, \label{e:opt_conv_power_balance} \\
\Pdc [m+1] &= P_{\mathrm{DER}}\,, \label{e:opt_dc_der} \\
\PlssC [i] &= k S_{c}[i] \: \forall \: i \in [1,\,m]\,, \label{e:opt_conv_loss} \\
\PlssN &= x^{\mathrm{T}} \Lambda x + \lambda x + \sigma \,, \label{e:opt_network_loss} \\
x &= [P_{c}^{\mathrm{T}},\,Q_{c}^{\mathrm{T}}]^{\mathrm{T}}\,, \label{e:opt_x_vector} \\
V &= Kx + b\,, \label{e:opt_voltage_mag} \\
V_{-} & \leq V \leq V_{+} \,, \label{e:opt_voltage_lim}\\
\sum _{i} S_{c}[i] &\leq  \ScTot \,, \label{e:opt_conv_sum}\\
\EC &\leq n \,. \label{e:opt_card_constraint}
\end{align}
The objective \eqref{e:opt_obj} is to schedule the converter real $ P_{c} $ and reactive $ Q_{c} $ power flows to minimize converter losses $ \PlssC $ and network losses $ \PlssN $. The apparent power of each converter $ S_{c}[i] $ must be bounded by the capacity connected to that converter leg \eqref{e:opt_apparent_limit}. The powers injected into the dc node $ \Pdc $ must balance \eqref{e:opt_dc_kcl}, and the power injected by a distributed energy resource $ P_{\mathrm{DER}} $ is the final element of the vector of powers at the dc node \eqref{e:opt_dc_der}. The power must balance across the lossy converters \eqref{e:opt_conv_power_balance}, with converter losses modelled as being linear in apparent power \eqref{e:opt_conv_loss} for loss coefficient $ k $. Network losses are quadratic in power injections  \eqref{e:opt_network_loss}, with $ x $ the stacked vector of real and reactive power injections \eqref{e:opt_x_vector} and $ \Lambda,\,\lambda,\,\sigma $ the parameters of the quadratic model. Voltage magnitudes $ V $ are linear in injection \eqref{e:opt_voltage_mag}, with $ K,\,b $ the sensitivity and offset components of the power-voltage affine model, and upper and lower bounds on voltages $ V_{+},\,V_{-} $ imposed via \eqref{e:opt_voltage_lim}. The total ac/dc converter capacity $ \ScTot $ limits the sum of power transfers into the ac feeders \eqref{e:opt_conv_sum}, based on the idealised MOP of \cite{deakin2023multiplexing_inpress}. Finally, the number of non-zero power injections (i.e., the electrical cardinality) is less than or equal to $ n $ \eqref{e:opt_card_constraint}.

Note that the full optimization formulation includes auxiliary variables to include the cardinality constraint \eqref{e:opt_card_constraint} as described in Section~\ref{ss:cardinality} (these are not written explicitly for brevity). Additionally, the quadratic network losses \eqref{e:opt_network_loss} are converted to a relaxed second order cone constraint, as in \cite{deakin2023multiplexing_inpress}.

\section{Case Studies}\label{s:results}

In the previous section, the operational optimization method was developed to optimally dispatch an idealised reconfigurable power converter, subject to a cardinality constraint that ensures the electrical cardinality is no greater than a given value. In this section, the approach is demonstrated on a pair of distribution network models, simulated over the course of a full calendar year.

\begin{figure}\centering
\subfloat[Case study 1: the 75 bus UKGDS HV UG network.]{\includegraphics[width=0.44\textwidth]{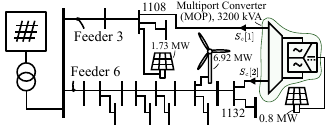}\label{f:case_study_hvug_sest23}}
~\\
\subfloat[Case study 2: the IEEE 33 Bus distribution system.]{\includegraphics[width=0.44\textwidth]{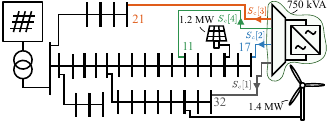}\label{f:case_study_33bus_sest23}}
\caption{The topology and generators installed for the two case studies, with the UKGDS having a three-port MOP with two ac feeders and a generator connected to the dc link (similar to Fig.~\ref{f:reconfig_design}); and, the 33 Bus system with a 4-terminal MOP.}
\label{f:case_study_ntwk_sest23}
\end{figure}

Fig.~\ref{f:case_study_ntwk_sest23} shows the topology of the interconnected feeders of the two networks studied, the 75 Bus UK Generic Distribution System (GDS) HV UG network \cite{foote2005ukgds} and the IEEE 33 Bus Network \cite{baran1989network}. Both of these models have a significant wind and solar generator connected within the network. Additionally, to illustrate how injections into the dc link of a multiport can influence the mission profile, the GDS network also has an 0.8~MW solar generator connected to the dc link. Each load is allocated an individual demand profile from real measured annual profiles from a utility in the UK \cite{deakin2023annual}, and embedded wind and solar profiles are taken from \cite{wilson2023espeni} to represent the renewable generation profiles. A converter loss coefficient $ k $ of 1\% is used (i.e., ac-dc-ac efficiency is 98\%), as in \cite{deakin2023multiplexing_inpress}.

The solution approach has been developed using the Mosek Fusion API \cite{mosek2021mosek}. The conic relaxation of \eqref{e:opt_network_loss} is accurate across all cases and time periods to within a relative error of $ 3\times 10^{-5} $. Relative and absolute integer optimization gaps of $ 10^{-4},\,10^{-5} $ are used, with all problems then solved to optimality. The quadratic loss model \eqref{e:opt_network_loss} has good performance as compared to the true non-linear solution obtained from OpenDSS. For example, for the 33 Bus system and unconstrained cardinality (Section~\ref{ss:case2}), the correlation between modelled and true loss changes has a correlation coefficient of 94.4\% and slope (as determined via linear regression) of 0.914. Linearization errors for voltage magnitudes are even more accurate--for example, relative errors across the whole year are less than 3.5\% for Case Study 1 (in terms of the 2-norm of the changes in voltages from the linear model and OpenDSS).

\subsection{Case Study 1: UKGDS Distribution System}\label{ss:case1}

\begin{figure}\centering
\subfloat[Power transfers, unrestricted cardinality ($ \EC $ unconstrained).]{\includegraphics[width=0.23\textwidth]{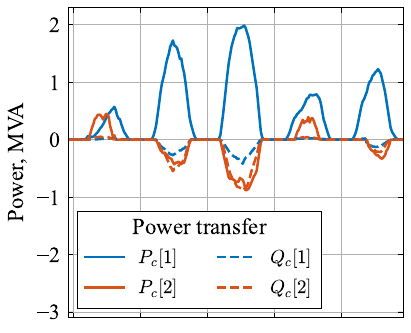}\label{f:pltCase1profiles_full}}~
\subfloat[Complex power transfer, cardinality constraint $ \EC\leq 1 $.]{\includegraphics[width=0.23\textwidth]{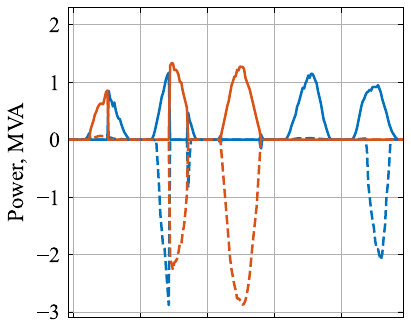}\label{f:pltCase1profiles_card}}\\
\subfloat[Additional system losses.]{\includegraphics[width=0.23\textwidth]{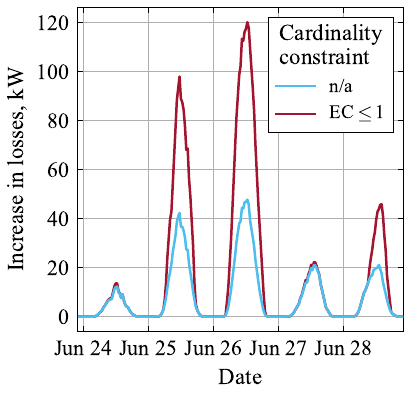}\label{f:pltCase1profile_losses}}~
\subfloat[Generation profiles and demand quartiles, Case 1.]{\includegraphics[width=0.23\textwidth]{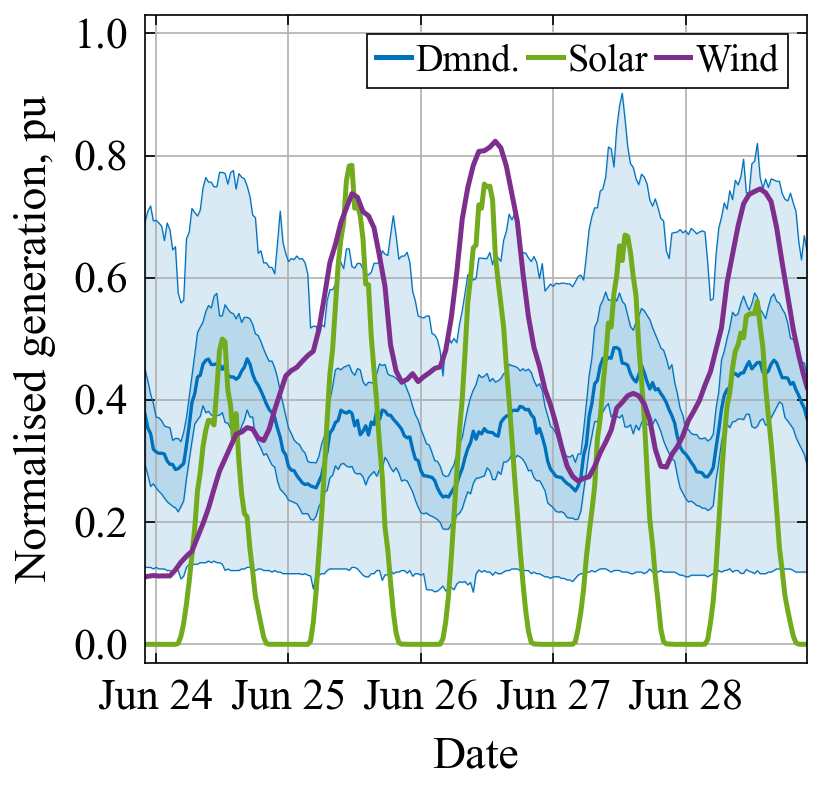}\label{f:pltCase1profiles_rnwbls}}\\
\caption{Power transfers (a, b), changes to losses (c) and generation profiles alongside the range, interquartile range and median demand (d) for Case 1 (UKGDS HVUG network, Fig.~\ref{f:case_study_hvug_sest23}). These subfigures show that power flow is feasible for the case with cardinality restricted to $ n =1$, but the additional reactive power that must be drawn to manage voltages increases losses substantially.}
\label{f:pltCase1profile}
\end{figure}

\begin{figure}\centering
\subfloat[Unrestricted cardinality.]{\includegraphics[width=0.23\textwidth]{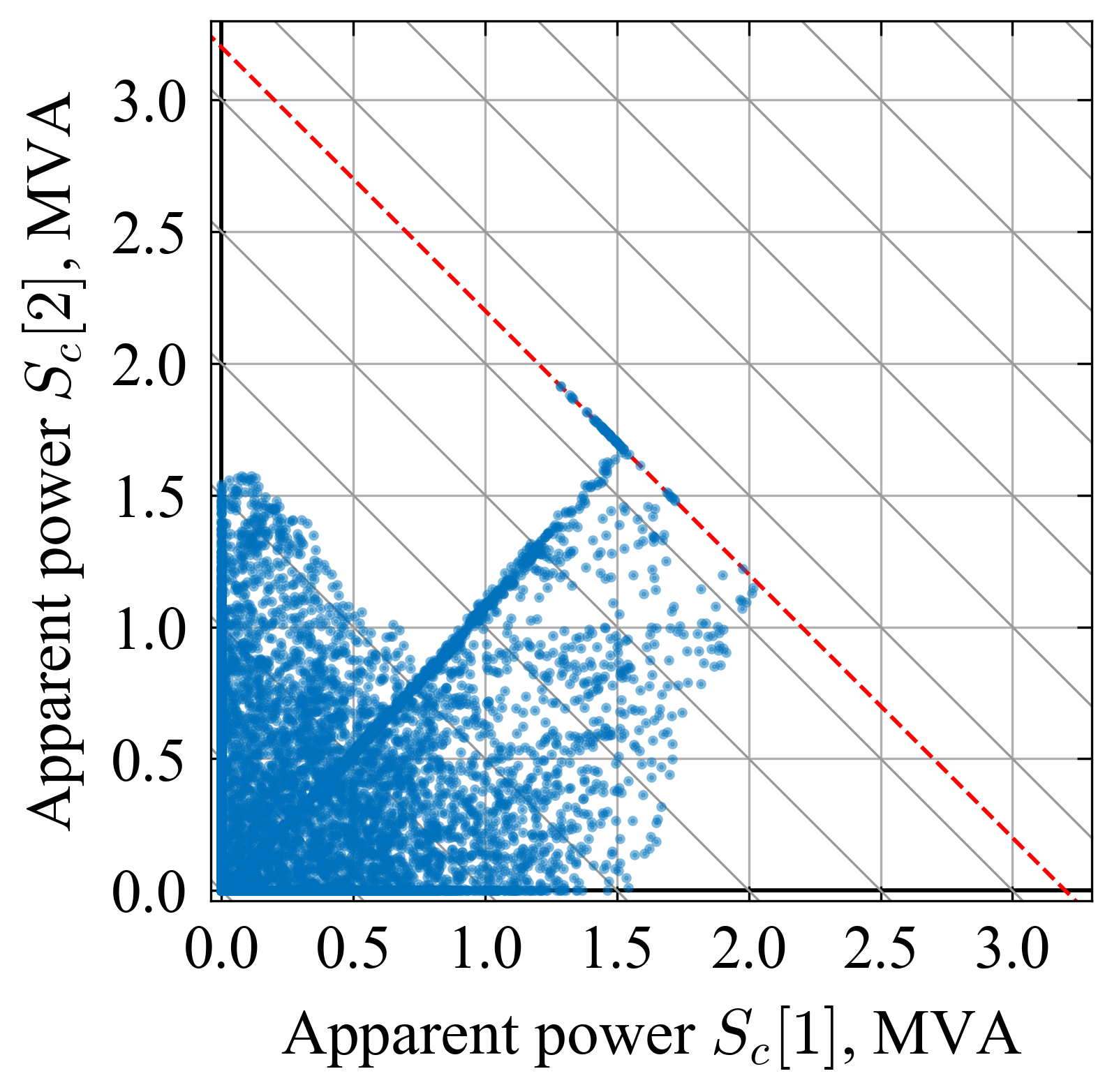}\label{f:pltCase1mp_full}}
~
\subfloat[Cardinality constraint $ \EC \leq 1 $.]{\includegraphics[width=0.23\textwidth]{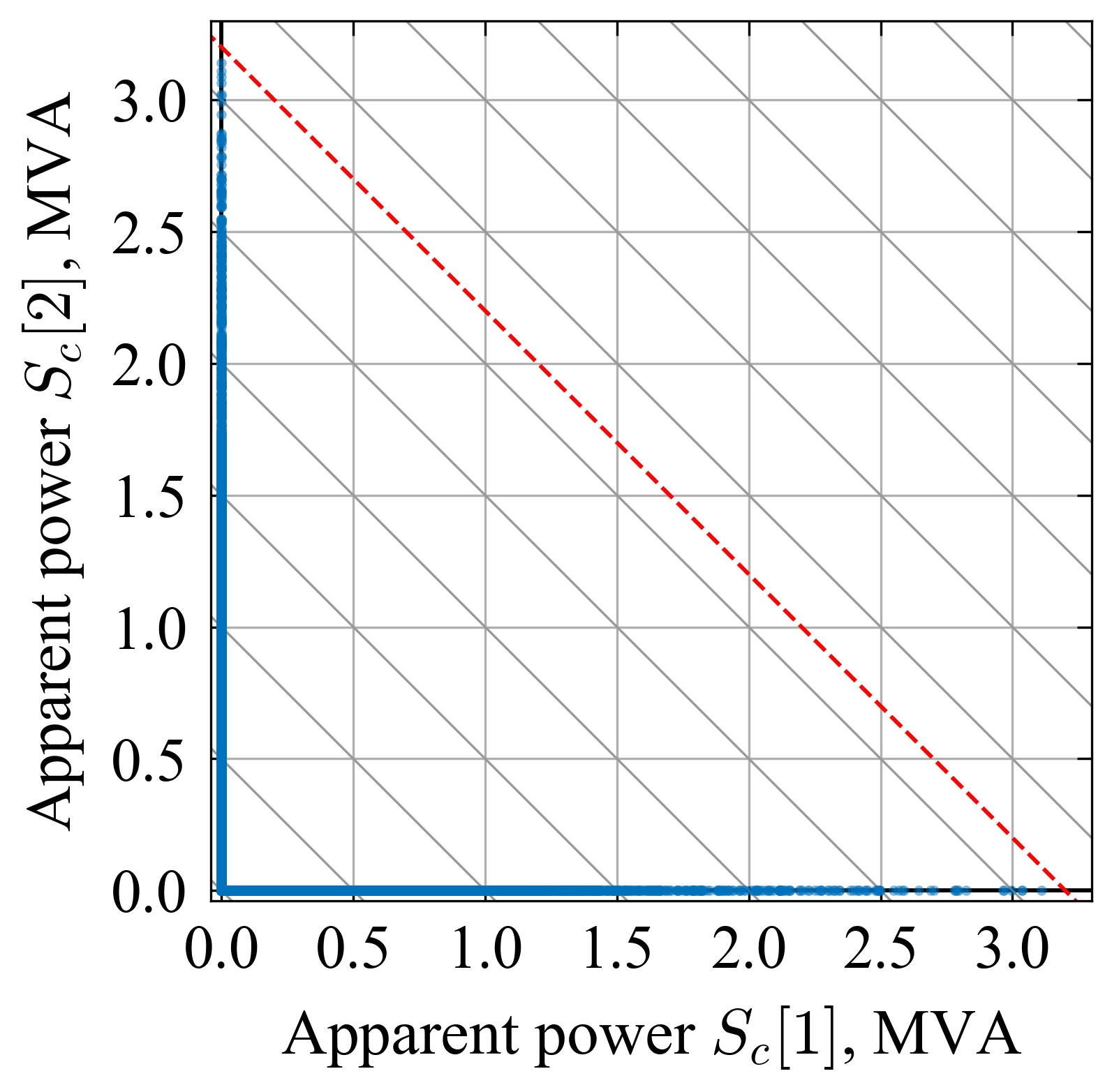}\label{f:pltCase1mp_card}}
\caption{Scatter plot of the mission profile for the unrestricted and constrained mission profiles, demonstrating cardinality EC of the mission profile has been reduced from 2 (a) to 1 (b). The thin red dashed line indicates the apparent power constraint for the 3200 kVA converter considered.}
\label{f:pltCase1mp}
\end{figure}

The first case considered is the UKGDS system (Fig.~\ref{f:case_study_hvug_sest23}), with total idealised converter capacity $ S_{c}^{\mathrm{Total}} $ of 3200~kVA. Fig.~\ref{f:pltCase1profile} plots the operation of the device for five representative days in the summer, showing the real and reactive flows of the two multiport converters with unrestricted cardinality (Fig.~\ref{f:pltCase1profiles_full}), the power flows for the converter with a cardinality constraint (Fig.~\ref{f:pltCase1profiles_card}), the additional losses in the system as compared to a system with no converter or the 0.8~MW additional generator (Fig.~\ref{f:pltCase1profile_losses}), and finally normalised values of the range and quartiles of the demand, the solar profile and wind profile (Fig.~\ref{f:pltCase1profiles_rnwbls}). It can be observed in this figure that, as expected, when the cardinality is unrestricted that power can be transferred between feeders in any combination. In contrast, when the cardinality EC is constrained, real and reactive powers can only be transferred into one feeder at a time (Fig.~\ref{f:pltCase1mp_full}). This has a substantial effect on power flows and losses, as there is voltage congestion at Bus 1132 (Fig.~\ref{f:pltCase1mp_full}) due to the large wind generator on Feeder 6 (for the purposes of this work, a limit of 1.045 pu is considered). For the restricted case with $ n=1 $, when there are high winds and solar generation (as on June 25th and 26th), the converter must connect to Feeder 6 to lower the voltage by drawing reactive power; however, this means that real power is also injected from the solar plant, and so further reactive power must be drawn to reduce the voltage even further. In contrast, the mission profile with unrestricted cardinality can simply transfer the surplus real power from the wind generator from Feeder 6 to Feeder~3.

Despite these increased system losses, congestion is still effectively managed by the power converter and so there are no infeasible points (i.e., time periods which the voltage is outside of the limits defined by the utility). Across the year, the increase in losses is 19.1 MWh for the converter with constrained cardinality as compared to the converter with unconstrained cardinality. However, this may be offset by the reduced complexity of the mission profile, as shown in Fig.~\ref{f:pltCase1mp}. A monolithic ac/dc converter with capacity 3200 kVA could be designed with a reconfiguration output stage to meet the mission profile profile, rather than a more complex ac/dc/ac converter.

\subsection{Case Study 2: IEEE 33 Bus Network}\label{ss:case2}

\begin{figure}\centering
\subfloat[Real power transfers, unrestricted cardinality ($ \EC $ unconstrained).]{\includegraphics[width=0.23\textwidth]{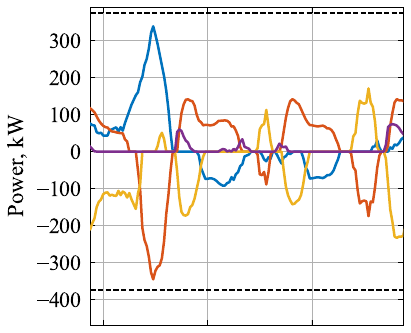}\label{f:pltSest2_powers_full}}~
\subfloat[Real power transfers, cardinality constraint $ \EC \leq 2 $.]{\includegraphics[width=0.23\textwidth]{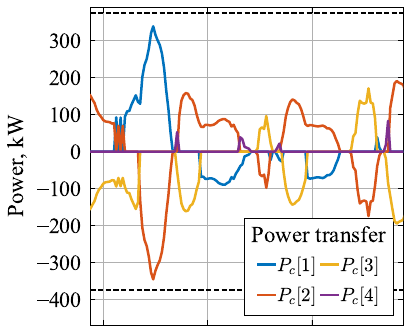}\label{f:pltSest2_powers_card2}}\\
\subfloat[Reduction in system losses.]{\includegraphics[width=0.23\textwidth]{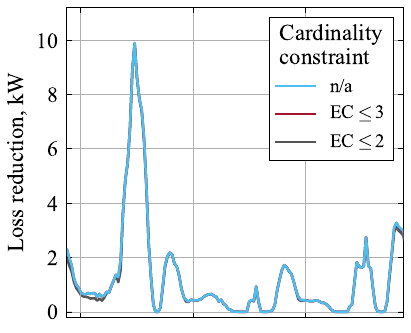}\label{f:pltSest2_actual_loss}}~
\subfloat[Fraction of reduction of system losses against cardinality constraint.]{\includegraphics[width=0.23\textwidth]{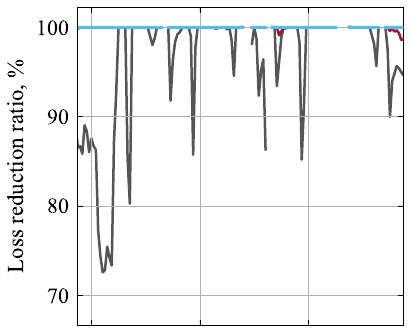}\label{f:pltSest2_loss_fraction}}\\
\subfloat[Number of non-zero apparent power transfers.]{\includegraphics[width=0.23\textwidth]{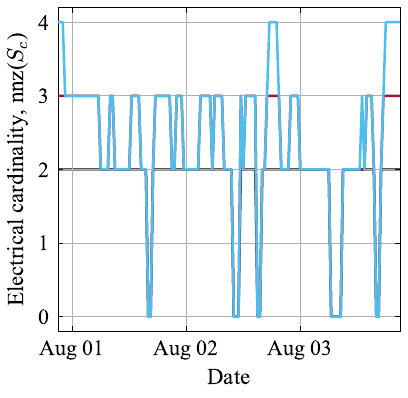}\label{f:pltSest2_nnz}}~
\subfloat[Generation profiles and demand quartiles, Case 2.]{\includegraphics[width=0.23\textwidth]{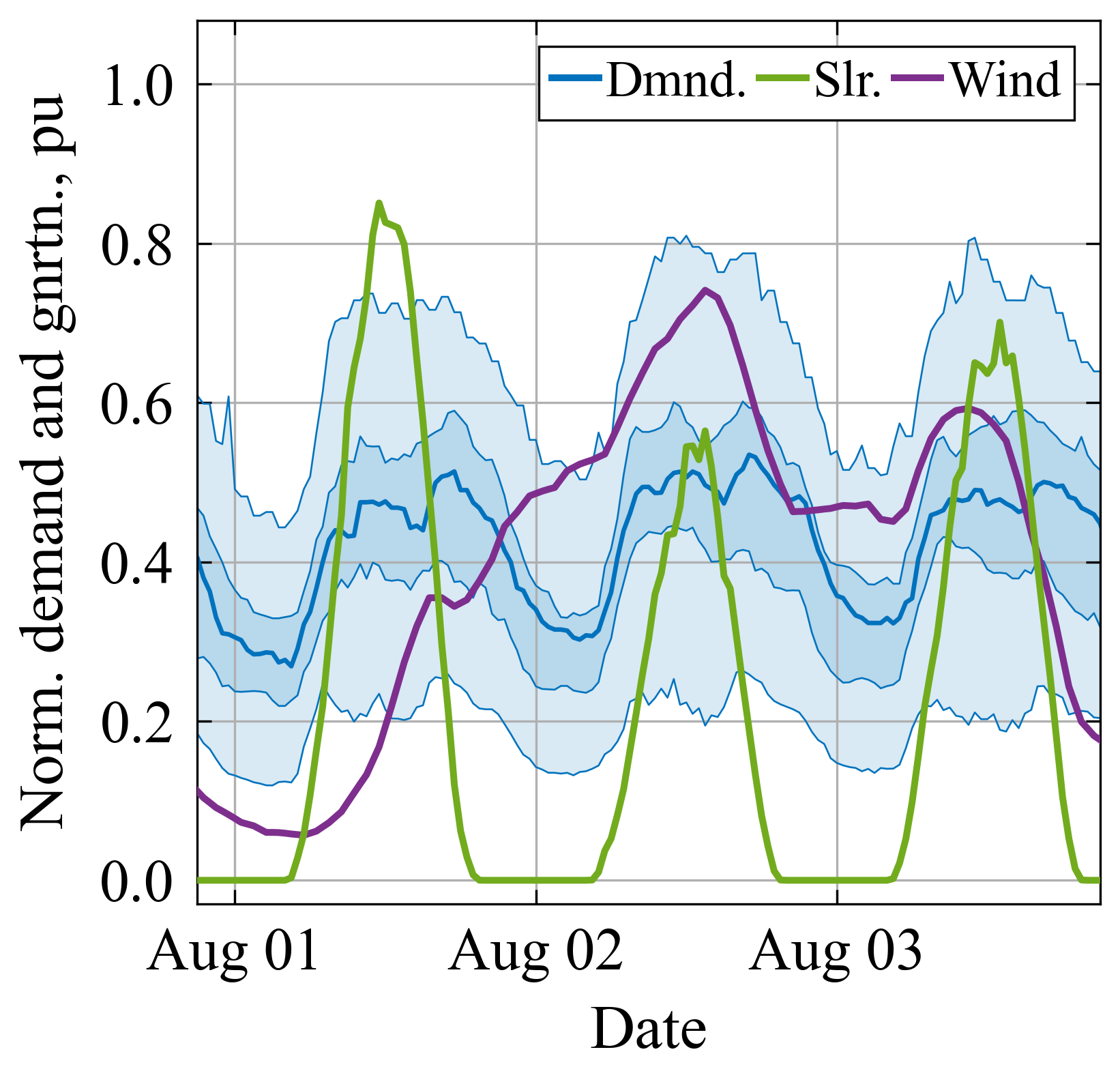}\label{f:pltSest2_demand_gen}}\\
\caption{Results for Case 2. (a, b) plot real power transfers for the unrestricted case ($ n=4 $) and for $ n=2 $; (c) plots the loss reduction, and (d) plots the equivalent fraction of loss reduction against the restricted cases. Finally, (e) plots the number of non-zero power transfers, whilst (f) plots the generation profiles and demand quartiles for this case.}
\label{f:pltSest2}
\end{figure}

The results with the IEEE 33 Bus system are plotted in Fig.~\ref{f:pltSest2} for representative days of August 1st-3rd, with a total ac/dc converter capacity $ S_{c}^{\mathrm{Total}} $ of 750 kVA. Real power transfers are plotted the unrestricted design in Fig.~\ref{f:pltSest2_powers_full} and the design with cardinality $ n $ restricted to 2 in Fig.~\ref{f:pltSest2_powers_card2}; the reduction in system losses are plotted in Fig.~\ref{f:pltSest2_actual_loss} and the relative fraction of the loss reduction as compared to the unrestricted cardinality are plotted in Fig.~\ref{f:pltSest2_loss_fraction} (to highlight more clearly the difference between the three lines in Fig.~\ref{f:pltSest2_actual_loss}); the out-turn cardinality of the hourly mission profile are plotted in Fig.~\ref{f:pltSest2_nnz}, and finally the normalised range and quartiles of the demand profiles, and the solar and wind profiles, are plotted in Fig.~\ref{f:pltSest2_demand_gen}. As expected, from this figure, it can again be seen that power can be transferred between any combination of powers for a converter with unrestricted cardinality, where the case with cardinality $ n $ restricted to 2 or less can only transfer power between any two feeders. For example, on August 1st, the high solar generation leads the MOP to draw from bus 17 $ P_{c}[2] $ and export to bus 32 $ P_{c}[1] $ due to high demand and low wind.

\begin{figure}\centering
\subfloat[Total annual losses]{\includegraphics[width=0.205\textwidth]{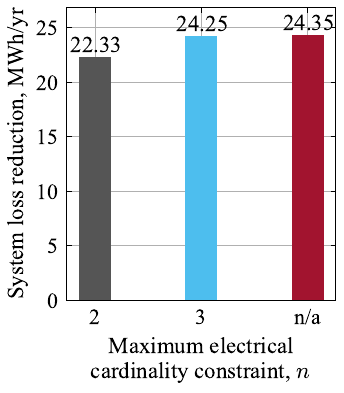}\label{f:plt_sest_case2_summary_loss}}
~
\subfloat[Distribution of EC values]{\includegraphics[width=0.24\textwidth]{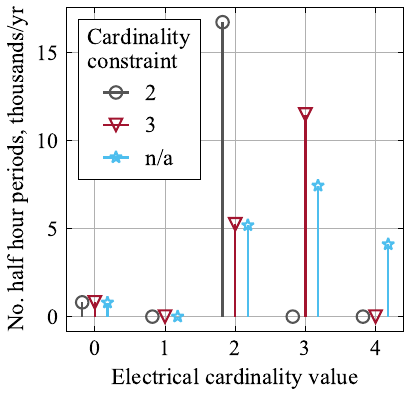}\label{f:plt_sest_case2_summary_nnz}}
\caption{Summary statistics for case study 2, in terms of (a) annual loss reduction and (b) the distribution of electrical cardinality (EC) values.}
\label{f:plt_sest_case2_summary}
\end{figure}

The results of Fig.~\ref{f:pltSest2} show clearly the fact that the cardinality EC can vary significantly as a function of time (calculated according to \eqref{e:nnz_tol}). As a result, during the time periods with an EC of 2, all three of the converter models perform the identically. Over the course of the full year (17520 half-hours), the losses for systems with 2, 3 or no cardinality constraint values are shown in Fig.~\ref{f:plt_sest_case2_summary}, as is the distribution of EC values. In this system, more than 4\% of time periods have no power injections (i.e., an EC value of zero), even with the low ac/dc conversion loss coefficient of $ 1 $\%. 

\subsection{Discussion}\label{ss:discussion}

This work has considered the construction of mission profiles with a constrained cardinality to explore the properties of these solutions. However, in general the idealised MOP model \cite{deakin2023multiplexing_inpress}, as used in the optimization \eqref{e:opt_conv_sum}, may or may not be realisable with fixed converter sizes (as modelled in, e.g., \cite{deakin2023multiplexing_inpress}). Nevertheless, the proposed approach can show if there are significant potential benefits of reduced cardinality, thereby motivating a future method to consider an optimal sizing strategy for individual ac/dc legs for a given cardinality.

Furthermore, this work has focussed on the \textit{electrical} cardinality, as this drives electrothermal stresses and is therefore the primary component of a mission profile. However, the thermal aspect of a mission profile has not been modelled explicitly, and the impact of changing cardinality may be considerable if a reconfigurable model is designed which substantially changes converter utilization. Conversely, distribution optimization problems could also be developed that aim to directly avoid unnecessary thermal cycling of power converter modules to improve reliability and lifetime of these devices.

\section{Conclusion}\label{s:conclusions}

Multiport mission profiles have more complex mathematical structure than two-terminal power converters, and so require more sophisticated techniques to model, design and simulate. The electrical cardinality of a multiport's mission profile has been introduced, based on the outputs of distribution system optimal power flow, and has been shown to have temporal variability within systems composed of power converters embedded within electrical distribution networks. Furthermore, a method to include cardinality constraints within optimal power flow problems has been proposed that enables a family of mission profiles to be provided to a power converter designer for further analysis. Results show that converters with restricted cardinality can still provide effective congestion management and provide grid services such as loss reduction. A case study demonstrating a halving of the cardinality only reducing the potential losses reduction by 91.7\% as compared to an unrestricted case.

Power flows in distribution systems are increasingly becoming controllable and variable, whilst developments in power electronics are resulting in increasing power density and reliability of converters. To maximise the potential benefits of these devices within systems, much higher levels of dialogue will be required between power converter designers and network operators. The electrical cardinality of the mission profile provides one useful piece of information that can be clearly communicated that affects both the power converter and its operation within a network. It is concluded that new integrated power distribution--power electronics design and modelling approaches will be necessary to most effectively realise the potential of power converters to support the integration of low carbon technologies in power distribution.

%

\bibliography{master_bib_egrid}{}

\begin{thebibliography}{10}
\providecommand{\url}[1]{#1}
\csname url@samestyle\endcsname
\providecommand{\newblock}{\relax}
\providecommand{\bibinfo}[2]{#2}
\providecommand{\BIBentrySTDinterwordspacing}{\spaceskip=0pt\relax}
\providecommand{\BIBentryALTinterwordstretchfactor}{4}
\providecommand{\BIBentryALTinterwordspacing}{\spaceskip=\fontdimen2\font plus
\BIBentryALTinterwordstretchfactor\fontdimen3\font minus
  \fontdimen4\font\relax}
\providecommand{\BIBforeignlanguage}[2]{{%
\expandafter\ifx\csname l@#1\endcsname\relax
\typeout{** WARNING: IEEEtran.bst: No hyphenation pattern has been}%
\typeout{** loaded for the language `#1'. Using the pattern for}%
\typeout{** the default language instead.}%
\else
\language=\csname l@#1\endcsname
\fi
#2}}
\providecommand{\BIBdecl}{\relax}
\BIBdecl

\bibitem{wang2019viable}
Z.~Wang, H.~Wang, Y.~Zhang, and F.~Blaabjerg, ``A viable mission profile
  emulator for power modules in modular multilevel converters,'' \emph{IEEE
  Transactions on Power Electronics}, vol.~34, no.~12, pp. 11\,580--11\,593,
  2019.

\bibitem{zhan2022multi}
C.~Zhan, L.~Zhu, W.~Wang, Y.~Zhang, S.~Ji, and F.~Iannuzzo, ``Multi-dimensional
  mission-profile-based lifetime estimation approach for {IGBT} modules in
  {MMC-HVDC} application considering bidirectional power transfer,'' \emph{IEEE
  Transactions on Industrial Electronics}, 2022.

\bibitem{sarantakos2022robust}
I.~Sarantakos, M.~Peker, N.~Zografou-Barredo, M.~Deakin, C.~Patsios,
  T.~Sayfutdinov, P.~Taylor, and D.~Greenwood, ``A robust mixed-integer convex
  model for optimal scheduling of integrated energy storage--soft open point
  devices,'' \emph{IEEE Transactions on Smart Grid}, vol.~13, no.~5, pp.
  4072--4087, 2022.

\bibitem{sun2020optimized}
F.~Sun, J.~Ma, M.~Yu, and W.~Wei, ``Optimized two-time scale robust dispatching
  method for the multi-terminal soft open point in unbalanced active
  distribution networks,'' \emph{IEEE Transactions on Sustainable Energy},
  vol.~12, no.~1, pp. 587--598, 2020.

\bibitem{lou2020new}
C.~Lou, J.~Yang, T.~Li, and E.~Vega-Fuentes, ``New phase-changing soft open
  point and impacts on optimising unbalanced power distribution networks,''
  \emph{IET Generation, Transmission \& Distribution}, vol.~14, no.~23, pp.
  5685--5696, 2020.

\bibitem{deakin2022design}
M.~Deakin, P.~C. Taylor, J.~Bialek, and W.~Ming, ``Design and operation of
  hybrid multi-terminal soft open points using feeder selector switches for
  flexible distribution system interconnection,'' \emph{Electric Power Systems
  Research}, vol. 212, no. 108516, 2022.

\bibitem{schaumburg2023efficiency}
J.~Schaumburg, J.~Kuprat, M.~Langwasser, and M.~Liserre, ``Efficiency
  optimization via mission profile-based power routing by design of hybrid grid
  connecting converter architecture,'' \emph{IEEE Open Journal of Power
  Electronics}, vol.~4, pp. 128--136, 2023.

\bibitem{deakin2023multiplexing_inpress}
M.~Deakin, ``Multiplexing power converters for cost-effective and flexible soft
  open points,'' \emph{IEEE Transactions on Smart Grid (In Press)}, 2023.

\bibitem{majumder2021distribution}
R.~Majumder, F.~Dijkhuizen, and B.~Berggren, ``Distribution networks with
  flexible direct current interconnection system,'' Apr.~6 2021, {US} Patent
  10,971,934.

\bibitem{jiang2022overview}
X.~Jiang, Y.~Zhou, W.~Ming, P.~Yang, and J.~Wu, ``An overview of soft open
  points in electricity distribution networks,'' \emph{IEEE Transactions on
  Smart Grid}, vol.~13, no.~3, pp. 1899--1910, 2022.

\bibitem{dall2014sparsity}
E.~Dall'Anese and G.~B. Giannakis, ``Sparsity-leveraging reconfiguration of
  smart distribution systems,'' \emph{IEEE Transactions on Power Delivery},
  vol.~29, no.~3, pp. 1417--1426, 2014.

\bibitem{boyd2009convex}
S.~Boyd and L.~Vandenberghe, \emph{Convex Optimization}, 1st~ed.\hskip 1em plus
  0.5em minus 0.4em\relax Cambridge University Press, 2009.

\bibitem{mosek2021cookbook}
Mosek, ``Mosek cookbook: Mixed integer optimization,''
  \url{https://docs.mosek.com/modeling-cookbook/mio.html\#maximum}, 2021,
  version 3.2.3. Accessed 2/10/21.

\bibitem{bernstein2018load}
A.~Bernstein, C.~Wang, E.~Dall’Anese, J.-Y. Le~Boudec, and C.~Zhao, ``Load
  flow in multiphase distribution networks: Existence, uniqueness,
  non-singularity and linear models,'' \emph{IEEE Transactions on Power
  Systems}, vol.~33, no.~6, pp. 5832--5843, 2018.

\bibitem{foote2005ukgds}
C.~Foote, P.~Djapic, G.~Ault, J.~Mutale, and G.~Strbac, ``United kingdom
  generic distribution system ({UKGDS}) - defining the generic networks,'' DTI
  Centre for Distributed Generation and Sustainable Electrical Energy, Tech.
  Rep., 2005.

\bibitem{baran1989network}
M.~E. Baran and F.~F. Wu, ``Network reconfiguration in distribution systems for
  loss reduction and load balancing,'' \emph{IEEE Power Engineering Review},
  vol.~9, no.~4, pp. 101--102, 1989.

\bibitem{deakin2023annual}
M.~Deakin, ``Annual half-hourly real and reactive power flows for 171 primary
  electrical distribution circuits in the north of england,'' DOI:
  10.25405/data.ncl.22047758, February 2023.

\bibitem{wilson2023espeni}
\BIBentryALTinterwordspacing
G.~Wilson, ``{Electrical half hourly raw and cleaned datasets for Great Britain
  from 2008-11-05},'' Mar. 2023. [Online]. Available:
  \url{https://doi.org/10.5281/zenodo.7740021}
\BIBentrySTDinterwordspacing

\bibitem{mosek2021mosek}
Mosek, ``Mosek fusion {API},'' \url{https://www.mosek.com/}, 2021.

\end{thebibliography}
\bibliographystyle{IEEEtran}

\end{document}